\documentclass[10pt,letterpaper]{article}
\usepackage{opex3}
\usepackage{amsmath,amsfonts,amssymb}
\usepackage{graphicx,cite}
\usepackage{color}
\pdfoutput=1
\begin{document}

\title{Optomechanically induced non-reciprocity in microring resonators}

\author{Mohammad Hafezi$^{1*}$ and Peter Rabl$^2$}

\address{$^1$Joint Quantum Institute, NIST/University of Maryland, College Park
20742, USA}
\address{$^2$Institute for Quantum Optics and Quantum Information of the Austrian
Academy of Sciences, 6020 Innsbruck, Austria}
\email{* hafezi@umd.edu}

\begin{abstract} We describe a new approach for on-chip optical non-reciprocity which makes use of strong optomechanical interaction  in microring resonators. By optically pumping the ring resonator in one direction, the optomechanical coupling is only enhanced in that direction, and consequently, the system exhibits a non-reciprocal response. For different configurations, this system can function either as an optical isolator or a coherent non-reciprocal phase shifter. We show that the operation of such a device on the level of single-photon could be achieved with existing technology.  
\end{abstract}


\section{Introduction}
The development of  integrated photonic circuits is a rapidly progressing field  which aims at the realization of micron scale photonic elements and the integration of these elements into a single chip-based device. Apart from conventional optical signal processing  and telecommunication applications \cite{photonics_book:2008},  this technology might  eventually also provide the basis for applications on  a more fundamental level such as optical quantum computation \cite{Matthews:2009,Obrien:2009p27731,Sansoni:2010,Politi:2009p45024} or photonic quantum simulation schemes \cite{Angelakis:2007,Greentree:2006,Hartmann:2006}.  A remaining challenge in integrated photonic circuits is on-chip optical isolation, that is, filtering of photons propagating in different directions in the circuit, or more generally, the implementation of non-reciprocal optical elements on a micrometer scale. 
Standard approaches for optical isolation  make use of
magneto-optical properties (e.g. Faraday rotation), which  however require
large magnetic fields \cite{Potton:2004}, and thus make it difficult
for integration \cite{Espinola:2004,Levy:2005,Zaman:2007} on a small scale.
To overcome this problem, other non-magnetic approaches have been proposed which, for example, rely on a dynamical modulation
of the index of refraction~\cite{Yu:2009}, stimulated inter-polarization scattering based on opto-acoustic effects \cite{Kang:2011}, modulated dielectric constant \cite{Feng:2011,Fan:2011} or on optical non-linearities that lead to an intensity
dependent isolation \cite{Soljacic:2003,Gallo:2001,Manipatruni:2009}.

The suitability of different optical isolation schemes  will depend very much on the specific task. While for many commercial applications  high bandwidth and robust fabrication techniques are key requirements for optical isolators, this can be different for on-chip quantum computing and quantum simulation schemes, where low losses, the operation on a single photon level and also the implementation of coherent non-reciprocal phase shifters are the most important aspects.   An intriguing new direction in this context is the study of  quantum
Hall physics with photons, which has recently attracted a lot of interest in the microwave as well as in the optical domain \cite{Koch:2010,Wang:2008,Wang:2009p16784,Haldane:2008,Hafezi:2011delay,Umucalilar:2011}. Here, apart from new possibilities to simulate quantum many body systems with light,  the appearance of edge states in quantum Hall system could also be exploited for a robust transfer of photons and optical delay lines. However, previous proposals (except Ref. \cite{Hafezi:2011delay}) can not be easily integrated
on chip,  while the scheme in Ref. \cite{Hafezi:2011delay} does not break the time reversal symmetry and therefore, it is not
suitable for non-reciprocal robust waveguides \cite{Wang:2009p16784}, or the emulation of  real magnetic fields for light.

In this work, we propose a new approach for  on-chip optical non-reciprocity which makes use of the recent advances in the fabrication of on-chip and micron sized optomechanical (OM)
devices \cite{Verhagen:2011,Chan:2011,Carmon:2007,Ding:2010}. In our scheme, the non-linear coupling between light and a mechanical mode inside a ring resonator leads to a non-reciprocal response of the OM system, which is  induced and fully controlled by an external driving field.  We characterize the input-output relations of such a  device and show that  by choosing different configurations  the same mechanism can be  employed  for  optical isolation as well as  non-reciprocal phase shifting and routing applications.  We describe under which conditions non-reciprocity is optimized and in particular,  we find that even in the presence of a finite intrinsic mode coupling inside the ring resonator, non-reciprocal effects remain large for a sufficiently strong OM coupling. In contrast to optical isolation based on a non-linear response of the OM system \cite{Manipatruni:2009}, our schemes can in principle be applied on a single photon level, limited by the up-conversion of thermal phonons only \cite{Stannigel:2010,Chang:2010,Stannigel:2011}. Our analysis shows that a noise level below a single photon can be achieved with present technology, which makes this device  a suitable building block for various non-reciprocal applications in the classical as well as the quantum regime.

\section{Optomechanically induced non-reciprocity: a toy model\label{sec:Toy-model}}

\begin{figure}
\center
\includegraphics[width=0.8\textwidth]{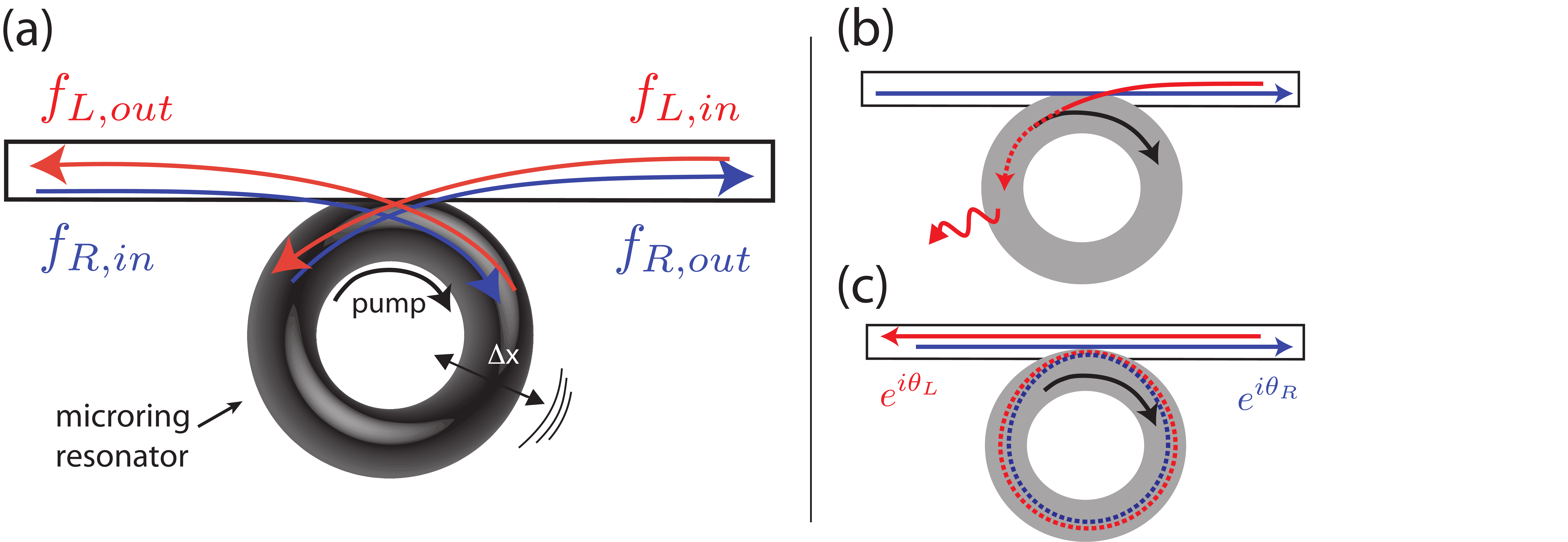}
\caption{Non-reciprocal optomechanical device. (a) A strong pump field enhances the optomechanical coupling between an isolated vibrational mode and the right-circulating optical mode inside a ring resonator. This results in different transmission properties for right- and left-moving fields in the waveguide. (b) Optical isolation.  (c) Non-reciprocal
phase shifter. \label{fig:input-output} }
\end{figure}

Before starting with a more general treatment below, we first outline in this section the essence of the OM induced
non-reciprocity for an idealized and slightly simplified setting. Specifically, we consider
an OM ring resonator, for example a toroidal microresonator, which is side-coupled to a waveguide as shown in Fig.~\ref{fig:input-output} (a). This configuration is 
commonly referred to as an all-pass filter (APF). The ring resonator supports two degenerate right- and left-circulating optical modes with frequency $\omega_c$ and bosonic operators $ a_R$ and $ a_L$ respectively. Radial vibrations of the resonator lead to a modulation of $\omega_c$  which can be modeled by the standard OM Hamiltonian \cite{Fabre:1994,WilsonRae:2007p27575,marquardt2007qtc,Schliesser:2010}
 $(\hbar=1)$, 
\begin{equation}
H_{om}=\omega_{m}  b^{\dagger}b+\sum_{i=L,R} \omega_c a_i^\dag a_i  +  g_0 a_{i}^{\dagger} a_{i}( b^{\dagger}+ b).
\end{equation}
Here $b$ is the bosonic operator for the mechanical mode of frequency $\omega_m$ and $g_0$ is the OM coupling, which corresponds to the optical frequency shift per quantum  of motion. Note that the mechanical mode is extended and varies slowly over the scale of the optical wavelength \cite{Schliesser:2010}. Therefore, the optomechanical coupling does not induce a mixing between the right- and left-circulating optical modes. In typical experiments $g_0$ is very weak and to enhance OM interactions we now assume that the right-circulating resonator mode is excited by an external laser field of frequency $\omega_L=\omega_c +\Delta$. In the limit $|\alpha_R|\gg1$, where $\alpha_R$ is the classical field amplitude of the driven mode, we can make a unitary transformation $ a_R\rightarrow  a_R + \alpha_R$  and linearize the OM coupling around $\alpha_R$. As a result, we obtain an effective Hamiltonian which in the frame rotating with $\omega_L$ is given by \cite{Fabre:1994,WilsonRae:2007p27575,marquardt2007qtc,Schliesser:2010}

\begin{equation}
H_{om-lin}  =  -\Delta(a_{R}^{\dagger}a_{R}+a_{L}^{\dagger}a_{L})+\omega_{m}b^{\dagger}b+(G_{R}a_{R}^{\dagger}+G_{R}^{*}a_{R})(b^{\dagger}+b), \label{eq:toy_hamiltonian}
\end{equation}
where an additional OM frequency shift has been reabsorbed into the definition of $\Delta$ (see Sec.~\ref{sec:Full-Model} for a more detailed derivation).   In Eq.~(\ref{eq:toy_hamiltonian}), we have introduced the \emph{enhanced} OM coupling $G_{R}=g_0 \alpha_R$, and in view of $|G_R|\gg g_0$, neglected residual OM interactions $\sim \mathcal{O}(g_0)$. We see that the external driving field creates  an asymmetry between left- and right-circulating modes, which we can exploit for generating non-reciprocal effects.

In order to study the transport properties of light through \textcolor{black}{such an opto-mechanical system}, we use the input-output
formalism \cite{Gardiner:1985,Fabre:1994}. For both propagation directions, we define in- and outgoing fields which are related by
\begin{equation}
f_{R/L,out}(t) =  f_{R/L,in}(t)  + \sqrt{2\kappa} a_{R/L}(t),
\end{equation}
where $2\kappa$ is the resonator decay rate into the waveguide.  \textcolor{black}{After Eq.~(\ref{eq:toy_hamiltonian}), the equations of motion takes the following form:
\begin{eqnarray}
\dot{ a_{R}} & = & (i\Delta-\kappa_{in}-\kappa) a_{R}-iG_R (b+b^\dagger) -\sqrt{2\kappa}f_{R,in},\\
\dot{ a_{L}} & = & (i\Delta-\kappa_{in}-\kappa) a_{L} -\sqrt{2\kappa}f_{L,in},\\
\dot{ b} & = & (-i\omega_{m}-\gamma_m) b-iG_R^*a_R-iG_Ra_R^\dagger,
\end{eqnarray}
where $2\kappa_{in}$ denotes the intrinsic photon loss rate of the optical resonator and $\gamma_m$ the mechanical damping rate. Due to the linearity of the above equations, we can solve them for the exception values, in frequency  space.} In the following, we are primarily interested in the case where the resonator is driven
at or close to the mechanical red sideband $(\Delta\approx -\omega_{m})$, and
only $a^{\dagger}b+ab^{\dagger}$ terms in Eq.~(\ref{eq:toy_hamiltonian}) will be resonant.  Therefore, in the sense of a rotating wave approximation, we can ignore other off-resonant contributions in Eq.~(\ref{eq:toy_hamiltonian}). In the appendix we show that this approximation is justified and it allows us to describe the transport properties of the systems in terms of a simple $2\times 2$  scattering matrix 
\begin{equation}\label{eq:scattering_matrix} 
\left(\begin{array}{c}
f_{R,out}(\delta)  \\
f_{L,out}(\delta)  \\
\end{array}\right)=
\left(\begin{array}{cc}
t_R(\delta) & r_L(\delta) \\
r_R(\delta) & t_L(\delta)  \\
\end{array}\right)
\left(\begin{array}{c}
f_{R,in}(\delta)  \\
f_{L,in}(\delta)  
\end{array}\right),
\end{equation}
where  $\delta=\omega+\Delta$ is the detuning of the incoming
field from the optical resonator resonance. For our idealized model, there is no scattering between left- and right-moving modes and $r_{R}=r_{L}=0$. In turn, the transmission coefficients are given by 
\begin{eqnarray}
t_{L}&=&\frac{\kappa_{in}-\kappa-i\delta}{\kappa_{in}+\kappa-i\delta}, \qquad t_{R}= 1-\frac{2(\gamma_{m}/2-i\delta)\kappa}{|G_{R}|^{2}+(\gamma_{m}/2-i\delta)(\kappa+\kappa_{in}-i\delta)}.
\label{eq:toy_scattering_coefs}\end{eqnarray}
 We see that the transmission spectrum of the left-going field is simply that of a bare \textcolor{black}{resonator}, while the transmission of the right-going mode  is modified by the presence of the mechanical oscillator.  In particular, the expression for $t_R$ resembles that of electromagnetically induced transparency (EIT)~\cite{fleischhauer1}, if we assume $G_{R}$
to be the ``control field''. As in atomic EIT, the coupling of the light field to the long-lived  mechanical mode ($\gamma_m\ll\kappa$) leads to a dip in the weak coupling limit and to a splitting of the transmission resonance in the strong coupling regime. This feature has already  been suggested in previous works for slowing and stopping of light using OM systems \cite{Agarwal:2010,Chang:2010,Weis:2010, SafaviNaeini:2011}. The same effect can be used for achieving non-reciprocity in the following ways:

\begin{figure}
\center \includegraphics[width=0.8\textwidth]{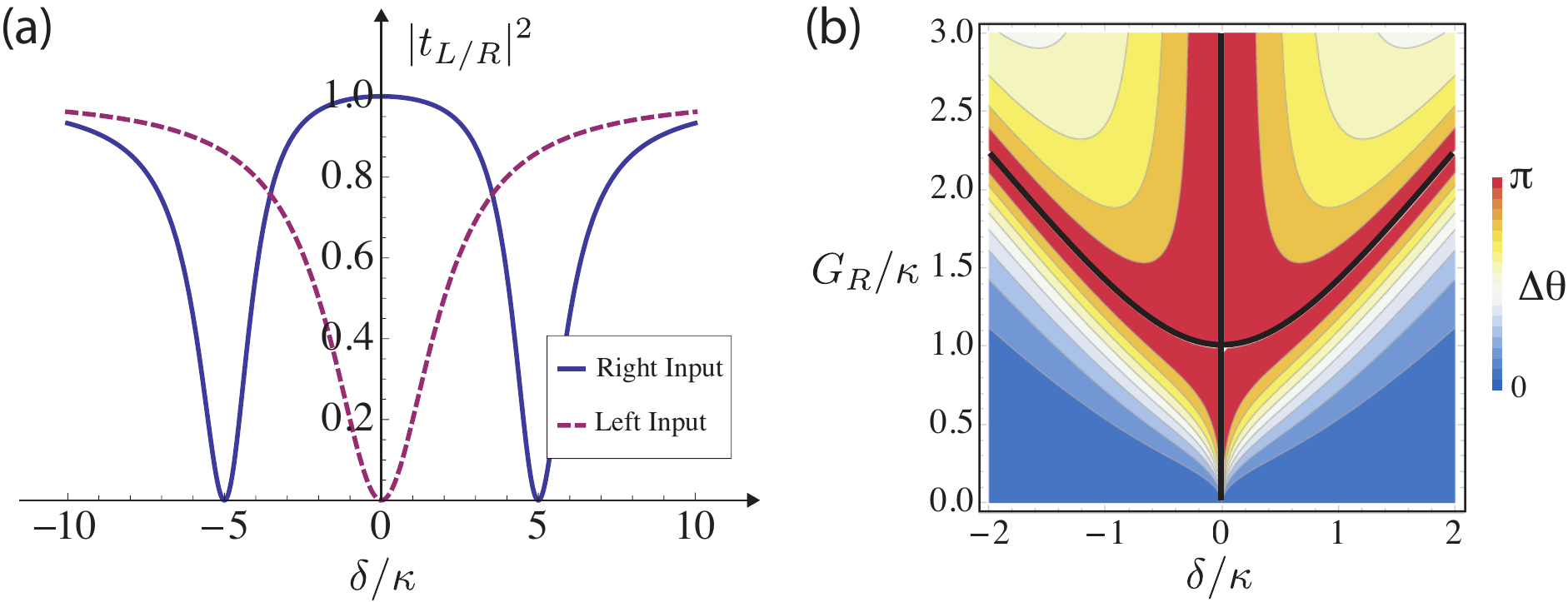}
\caption{(a) Transmission $|t_{R/L}|^2$ of the OM system when operated as an optical isolator ($\kappa_{in}=\kappa$). Within the resonator
bandwidth, the left-moving field is attenuated while the right-moving field is almost completely
transmitted. For this plot $G_R=5\kappa$. (b) Non-reciprocal phase shifter ($\kappa_{in}=0.01\kappa$). Both the left and the right input field are almost completely transmitted ($>98\%$), but 
acquire different phases, $\Delta \theta=\theta_R-\theta_L$. Black lines show the location of resonances. For these plots $\gamma_m=0$. \label{fig:toy_model}}
\end{figure}

\emph{Optical isolation.}  Let us first consider a critically coupled ring resonator where $\kappa=\kappa_{in}$. In this case, we see that for frequencies around the ring optical resonance ($\delta \approx 0$),
\begin{equation}
\left(\begin{array}{c}
f_{R,out}(\delta)  \\
f_{L,out}(\delta)  \\
\end{array}\right)\approx 
\left(\begin{array}{cc}
1 & 0 \\
0 & 0  \\
\end{array}\right)
\left(\begin{array}{c}
f_{R,in}(\delta)  \\
f_{L,in}(\delta)  \\
\end{array}\right). 
\end{equation}
Therefore, this configuration realizes an optical diode,  where light passes unaltered in one direction, but is completely absorbed in the other direction, as schematically
shown in Fig.~\ref{fig:input-output}(b). The frequency window over which this isolation is efficient is approximately $G_{R}^{2}/\kappa$ in the weak
coupling limit $(G_{R}\ll\kappa)$ and $\kappa$ in the strong coupling limit, where  the width of the EIT window is $2G_{R}$ and exceeds the resonator linewidth. A typical non-reciprocal transmission spectrum for the strong coupling regime is shown in Fig.~\ref{fig:toy_model}(a), which is that of an optical diode for frequencies around $\delta \approx 0$. Note that in contrast
to conventional optical isolation, no magnetic field is applied and instead the optical pump breaks the left-right symmetry.

\emph{Non-reciprocal phase shifter.}  
Let us now consider the so-called over-coupled regime where the intrinsic loss is much smaller than the resonator-waveguide coupling ($\kappa_{in}\ll\kappa$). In this case, the transmittance
is close to unity in both directions. However, the left- and right-going fields experience a different dispersion and 
\begin{equation}
\left(\begin{array}{c}
f_{R,out}(\delta)  \\
f_{L,out}(\delta)  \\
\end{array}\right)\approx 
\left(\begin{array}{cc}
e^{i\theta_R(\delta)}  & 0 \\
0 & e^{i\theta_L(\delta)} \\
\end{array}\right)
\left(\begin{array}{c}
f_{R,in}(\delta)  \\
f_{L,in}(\delta)  \\
\end{array}\right). 
\end{equation}
In general, the phases $\theta_R$ and $\theta_L$ will be different, and therefore, in this configuration, our devices acts as a non-reciprocal phase shifter,  as schematically shown in Fig. \ref{fig:input-output}(c). Again, in contrast to conventional magnetic field induced non-reciprocal phases, e.g. Faraday rotation, our scheme does not require large magnetic fields. As shown in Fig. \ref{fig:toy_model}(b), the OM induced phase difference $|\theta_R-\theta_L|$ can easily be controlled by changing the pump intensity and can be tuned from zero to about $\pi$ over a large range of frequencies. Therefore, a maximal non-reciprocal phase shift can already be achieved for light passing through a single device.

\section{General formalism\label{sec:Full-Model}}

\begin{figure}
\center
\includegraphics[width=0.4\textwidth]{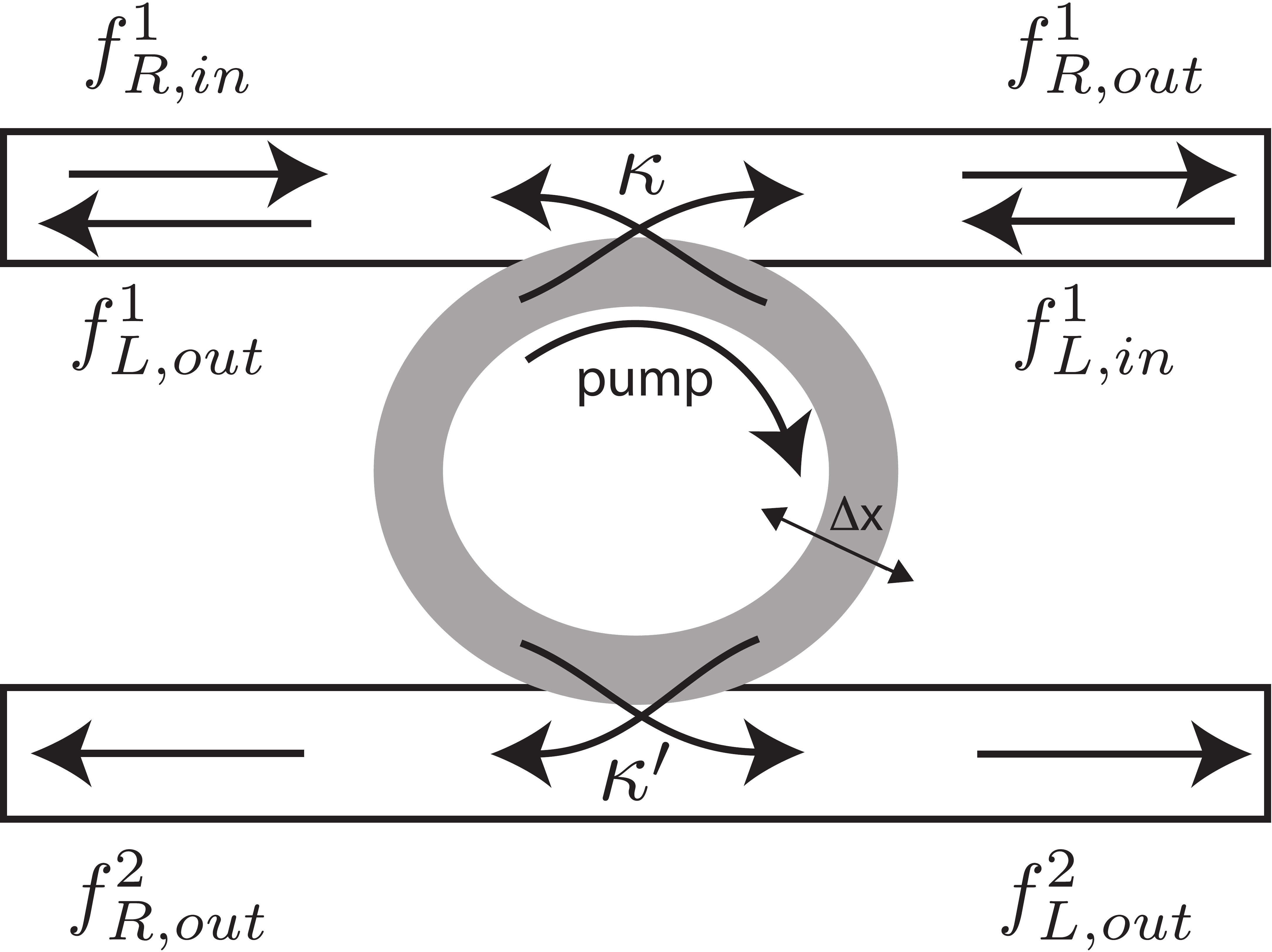}
\caption{General add-drop configuration, which can be employed for non-reciprocal photon routing between the upper and lower waveguide. It reduces to \textcolor{black}{the resonator coupled to a single waveguide}, if 
the coupling to the lower waveguide is absent ($\kappa'=0$).}
\label{fig:add-drop}
\end{figure}

In this section we present the general formalism for investigating
OM induced non-reciprocity. In particular, we now include
the effect of energy non-conserving terms as well as a finite coupling between the left- and right-circulating resonator modes which have been neglected
in our simplified discussion above. For completeness, we will also extend our discussion to a slightly more general configuration shown in Fig. \ref{fig:add-drop}, where the ring resonator is side-coupled to two optical waveguides with rates $\kappa$ and $\kappa'$. For $\kappa'=0$, this setting reduces to \textcolor{black}{the resonator coupled to a single waveguide case, which was discussed above}. Moreover, in the so-called add-drop configuration ($\kappa'=\kappa$, $\kappa_{in}\approx 0$), this device can be used for non-reciprocal routing of light.

To account for a more realistic situation we now include the presence of intrinsic defects inside the ring resonator and model the system by the total Hamiltonian \begin{equation}
H= H_{om} + \beta a_{L}^{\dagger}a_{R}+ \beta^* a_{R}^{\dagger}a_{L}.
\end{equation}
Here, in addition to the OM interaction $H_{om}$ given in Eq.~(\ref{eq:toy_hamiltonian}),  the second and third terms in this Hamiltonian represent a coherent scatting of strength $\beta$ between the two degenerate optical modes, which  is associated with bulk or surface imperfections \cite{Kippenberg:2002,Mazzei:2007}. The system dynamics is fully described by the set of quantum Langevin equations $(i=L,R)$ 
\begin{eqnarray}
\dot{a_{i}}(t) & = & i[H,a_{i}(t)]-\kappa_t a_{i}(t)-\sqrt{2\kappa}f^1_{i,in}(t)-\sqrt{2\kappa'}f^2_{i,in}(t)- \sqrt{2\kappa_{in}} f_{i,0}(t),\\
\dot{b}(t) & = & i[H,b(t)]-\frac{\gamma_{m}}{2}b(t)-\sqrt{\gamma_{m}}\xi(t) \label{eq:heisenberg_mech},
\end{eqnarray}
together with the relations $f^1_{i,out}(t)  =  f^1_{i,in}(t)+\sqrt{2\kappa}a_{i}(t)$ and $f^2_{i,out}(t)  =  f^2_{i,in}(t)+\sqrt{2\kappa'}a_{i}(t)$ between the in- and out-fields.  
In these equations,  $\kappa_t=\kappa+\kappa'+\kappa_{in}$ is the total ring resonator field decay rate and  
the operators $f_{i,in}^{1,2}(t)$ are $\delta$-correlated field operators for the in-fields in the upper and lower waveguide (see Fig.~\ref{fig:add-drop}) and $f_{i,0}(t)$ is a vacuum noise operator associated with the intrinsic photon loss.  
Finally, $\gamma_m=\omega_m/Q_m$ is the mechanical damping rate for a quality factor $Q_m$ and $\xi(t)$ is the corresponding noise operator.  In contrast to the optical fields, the mechanical mode is coupled to a reservoir of finite temperature $T$ such that $[\xi(t), \xi^\dag(t')]   =\delta(t-t')$ and $\langle \xi(t) \xi^\dag(t')\rangle = (N_{th}+1)\delta(t-t')$ where $N_{th}$ is the thermal equilibrium occupation number of the mechanical mode. Note that the Langevin equation for the mechanical mode, Eq.~(\ref{eq:heisenberg_mech}), is only valid for $\gamma_m\ll \omega_m$. For typical mechanical quality factors $Q_m\sim 10^4-10^5$ this condition is well satisfied and for most of the results discussed below we will consider the limit $\gamma_m\rightarrow 0$, while keeping a finite thermal heating rate $\gamma_ m N_{th}\simeq k_BT/(\hbar Q_m)$.

As before, we assume that the clockwise mode of the
resonator is driven by a strong classical field of frequency $\omega_{L}=\omega_{c}+\Delta_{0}$
and amplitude $\mathcal{E}.$ We make the transformation $f_{R,in}(t)\rightarrow f_{R,in}(t)+\sqrt{2\kappa}\mathcal{E}$
and write the average field expectation values in the frame rotating with $\omega_L$, 
\begin{eqnarray}
\dot{\langle a_{R}\rangle} & = & (i\Delta_{0}-ig_0\langle b+b^{\dagger}\rangle-\kappa_t)\langle a_{R}\rangle-i\beta^*\langle a_{L}\rangle-2\kappa\mathcal{E},\\
\dot{\langle a_{L}\rangle} & = & (i\Delta_{0}-ig_0\langle b+b^{\dagger}\rangle-\kappa_t)\langle a_{L}\rangle-i\beta\langle a_{R}\rangle,\\
\dot{\langle b\rangle} & = & -i\omega_{m}\langle b\rangle-ig_0(|\langle a_{R}\rangle|^{2}+|\langle a_{L}\rangle|^{2}).
\end{eqnarray}
In the steady-state, we find that $\langle b\rangle=-g_0(|\langle a_{R}\rangle|^{2}+|\langle a_{L}\rangle|^{2})/\omega_{m}$.
By redefining the detuning to absorb the OM shift, $\Delta=\Delta_{0}+2g^{2}_0(|\langle a_{R}\rangle|^{2}+|\langle a_{L}\rangle|^{2})/\omega_{m}$,
we can rewrite the optical field equations in the steady state as
\begin{eqnarray}
0 & = & (i\Delta-\kappa_t)\langle a_{R}\rangle-i\beta^*\langle a_{L}\rangle-2\kappa\mathcal{E},\\
0 & = & (i\Delta-\kappa_t)\langle a_{L}\rangle-i\beta\langle a_{R}\rangle.\end{eqnarray}
 In the absence of mode coupling ($\beta=0$$)$, the counter clockwise
mode remains empty $(\langle a_{L}\rangle=0$), and we obtain $\langle a_{R}\rangle=2\kappa\mathcal{E}/(i\Delta-\kappa_t)$.
However, in the presence of mode coupling, we have
\begin{equation}
\langle a_{R}\rangle=\frac{2\kappa(i\Delta-\kappa_t)}{(i\Delta-\kappa_t)^{2}+|\beta|^{2}}\mathcal{E},\qquad \langle a_{L}\rangle=\frac{i\beta}{i\Delta-\kappa_t}\langle a_{R}\rangle, \label{eq:pump_inside}\end{equation}
 and in general both optical modes are excited. As above, we proceed by making the unitary transformations $a_{i}\rightarrow a_{i}+\langle a_{i}\rangle$
and $b\rightarrow b+\langle b\rangle$ and after neglecting terms of $\mathcal{O}(g_0)$,
we arrive at the linearized OM Hamiltonian
\begin{eqnarray}\label{eq:hamiltonian}
H  &= &  \omega_{m}b^{\dagger}b - \sum_{i=R,L} \Delta a_{i}^{\dagger}a_{i} +\beta a_{L}^{\dagger}a_{R}+\beta^*a_{R}^{\dagger}a_{L}\\ \nonumber
&&+ \sum_{i=R,L} (G_{i}a_{i}^{\dagger}+G_{i}^{*}a_{i})(b^{\dagger}+b),
\end{eqnarray}
where due to the mode coupling, both circulating modes exhibit an enhanced coupling ($G_i=g_0 \alpha_i$) to the mechanical mode. We are primarily interested in
the case where the resonator is driven near the mechanical red sideband
$(\Delta=-\omega_{m})$, where the terms of the form $a^{\dagger}_ib+a_i b^{\dagger}$
are dominant. However, small corrections due to the off-resonant couplings $a_i^{\dagger}b^{\dagger}+a_i b$
are included in our general formalism.

We group the OM field operators into a vector $v(t)=(b(t),a_{R}(t),a_{L}(t),b^{\dagger}(t),a_{R}^{\dagger}(t),a_{L}^{\dagger}(t))^{T}$
and write the equations of  motion in the form
\begin{equation}
\partial_{t}v(t)=-Mv(t)-\sqrt{2\kappa}I^{1}(t)-\sqrt{2\kappa'}I^{2}(t)- \sqrt{\gamma_m} I^{m}(t).\label{eq:noise}\end{equation}
Here the coupling matrix $M$ is given by 
\begin{equation}
M=i\left(\begin{array}{cccccc}
\omega_{m}-i \gamma_m/2 & G_{R}^{*} & G_{L}^{*} & 0 & G_{R} & G_{L}\\
G_{R} & -\Delta-i\kappa_t & \beta^{*} & G_{R} & 0 & 0\\
G_{L} & \beta & -\Delta-i\kappa_t & G_{L} & 0 & 0\\
0 & -G_{R}^{*} & -G_{L}^{*} & -\omega_{m}-i \gamma_m/2 & -G_{R} & -G_{L}\\
-G_{R}^{*} & 0 & 0 & -G_{R}^{*} & \Delta-i\kappa_t& -\beta\\
-G_{L}^{*} & 0 & 0 & -G_{L}^{*} & -\beta^{*} & \Delta-i\kappa_t \end{array}\right),\end{equation}
and the input field vectors are defined as $I^{i}(t)=(0,f_{R,in}^{i}(t),f_{L,in}^{i}(t),0,f_{R,in}^{\dagger i}(t),f_{L,in}^{\dagger i}(t))^T$ for $i=1,2$ and
$I^{m}(t)=(\xi(t),0,0,\xi^\dag(t),0,0)^T$.
Note that in Eq.~(\ref{eq:noise}), we have already omitted contributions from the intrinsic noise operators $f_{i,0}(t)$ which act on the vacuum and therefore do not contribute to the results discussed  below.
We solve the equations of motion for the OM degrees of freedom
($a$'s and $b$'s) in the Fourier domain and obtain
\begin{equation}
\tilde{v}(\omega)  =  (-M+i\omega\mathbb{I})^{-1}(\sqrt{2\kappa}\tilde{I}^{1}(\omega)+\sqrt{2\kappa'}\tilde{I}^{2}(\omega)+\sqrt{\gamma_m} \tilde I^{m}(\omega)).\label{eq:full_fourier}\end{equation}
By defining the output field vector
\begin{equation}
\tilde{O}^{1}(\omega)=(0,\tilde{f}_{R,out}^{1}(\omega),\tilde{f}_{L,out}^{1}(\omega),0,\tilde{f}_{R,out}^{\dagger1}(-\omega),\tilde{f}_{L,out}^{\dagger1}(-\omega))^T,
\end{equation}
we can rewrite the input-output relation as
\begin{equation}
\tilde{O}^{1}(\omega)=\sqrt{2\kappa}\, diag(0,1,1,0,1,1)\tilde{v}(\omega)+\tilde{I}^{1}(\omega),\label{eq:full_output}\end{equation}
and a similar expression can be derived for the out-fields in the second waveguide $\tilde{O}^{2}(\omega)$. Combining Eqs. (\ref{eq:full_fourier}-\ref{eq:full_output}),
the output fields can be evaluated as a function of the input field for arbitrary system parameters. 
Note that due to the presence of non-resonant OM interactions $\sim (a_i^\dag b^\dag + a_i b)$,  the scattering matrix mixes the $f_{i,in}$ with the conjugate fields $f_{i,in}^\dag$. In other words this means that different quadratures of the input fields have different transmission properties, an effect which is related to OM squeezing \cite{Fabre:1994,Mancini:1994,Brooks:2011}. However, in the appendix we show that for the relevant parameter regimes this effect is negligible in our device and for a more transparent discussion we will evaluate below only the relevant phase independent part of the scattering matrix.

\section{Results and discussion}

In the four port device shown in Fig. \ref{fig:add-drop}, we can study various different non-reciprocal effects and apart from the optical diode and phase-shifter settings outlined above the add-drop configuration ($\kappa=\kappa'$, $\kappa_{in}=0$) could be used to realize a non-reciprocal optical router between the two waveguides where, e.g.,  $f^1_{R,in}\rightarrow f^1_{R,out}$ but $f^1_{L,in}\rightarrow f^2_{L,out}$. However, this situation is formally equivalent to the optical diode by interchanging the role of $\kappa'$ and $\kappa_{in}$ and therefore we can restrict the following discussion to the transmission amplitudes $t_{R,L}(\omega)$ as defined in the two port scattering matrix in Eq.~(\ref{eq:scattering_matrix}).

\begin{figure}
\center 
\includegraphics[width=0.4\textwidth]{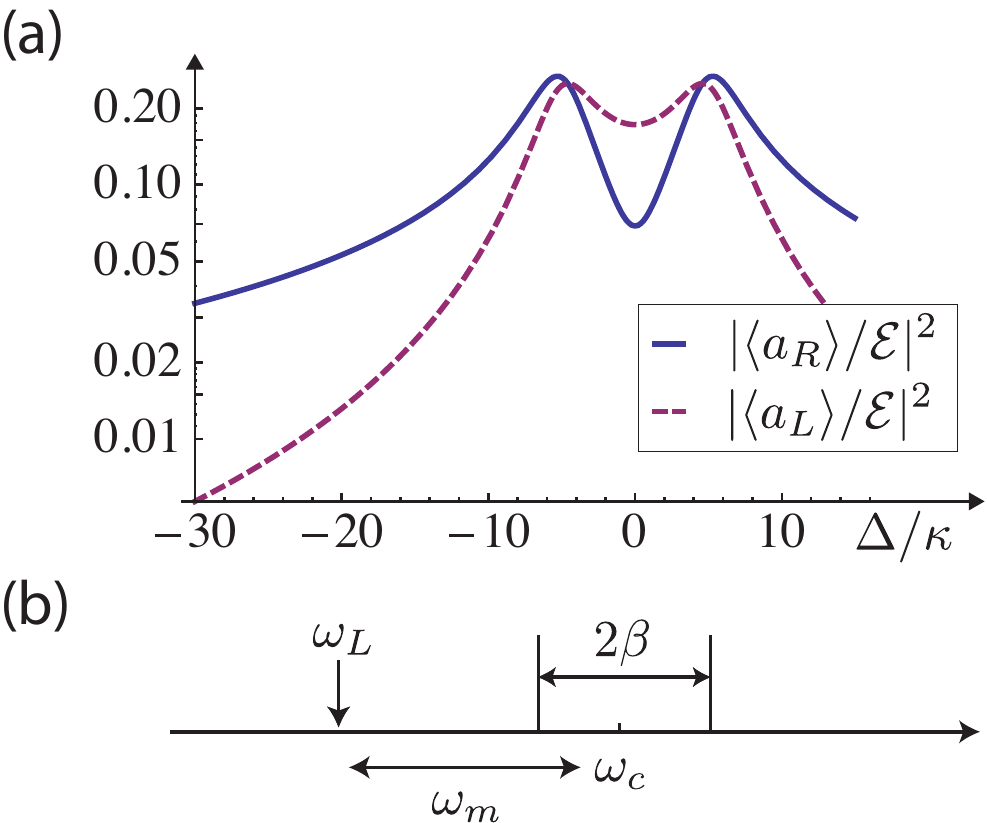}
\caption{Mean photon number in the left and right circulating modes in the presence of a finite mode coupling $\beta$ and as a function of the pump detuning $\Delta=\omega_L-\omega_c$. For this plot we have assumed that the pump field only drives the right-circulating mode and that the resonator is
coupled to a single waveguide ($\kappa'=0$).  The other parameters are $(\beta,\kappa_{in})/\kappa=(4,1)$. At the normal mode frequencies $\omega\simeq\pm\beta$,
the left- and right-circulating modes are almost equally populated, while everywhere else, there is an intensity imbalance between left- and right-circulating
modes. (b) The diagram shows the relation between the relevant frequencies in the system. In the presence of the mode coupling, the sidebands
($\pm\beta$) are located around the bare resonator frequency $\omega_{c}$ and the resonator is pumped at the mechanical red sideband. }
\label{fig:mode-coupling}
\end{figure}

Compared to the ideal situation described in Sec. \ref{sec:Toy-model}, we are now in particular interested in OM non-reciprocity in the presence of a finite intrinsic mode coupling, $\beta\neq 0$, where photons in the left- and right-circulating modes of both the
probe and pump field can no longer  propagate independently. Such a coupling is found in many experiments with high-Q micro-resonators and often attributed to bulk or surface imperfections \cite{Kippenberg:2002,Mazzei:2007}.  As  already mentioned above, a first consequence of this mode mixing is that the pump field is scattered into the left-circulating mode and we obtain enhanced OM couplings $G_{R,L}\sim \alpha_{R,L}$ for both propagation directions (see Eq.~(\ref{eq:pump_inside})). More specifically, for a purely right-going pump field, the intra-resonator fields are given by
\begin{eqnarray}
\frac{\langle \alpha_R\rangle}{\mathcal{E}} &=&\frac{2\kappa(i\Delta-\kappa_t)}{(i\Delta-\kappa_t)^{2}+|\beta|^{2}},\qquad \frac{\langle\alpha_L\rangle}{\mathcal{E}} =\frac{2i\beta\kappa}{|\beta|^{2}+(i\Delta-\kappa_t)^{2}},\end{eqnarray}
and these expressions are plotted in Fig. \ref{fig:mode-coupling} as a function of the pump detuning $\Delta$ and for the case of large mode coupling ($\beta\gg\kappa_t$). We see that in principle an asymmetric pumping can be achieved either for $\Delta=0$ or $|\Delta| \gg \beta$. However, to achieve a resonant  OM coupling, we should choose $\Delta\simeq -\omega_m$. Therefore,  $|G_L|/|G_R| \sim  \beta/\sqrt{\omega_m^2+\kappa^2}$ which means that the parasitic coupling can be suppressed by choosing high frequency mechanical modes. Further, we point out that a complete cancellation of $G_L$  could be achieved by adding a second pump beam in the left-circulating direction.  In particular, if the strength of the left input pump is chosen as $\mathcal{E'}=-i\beta/(i\Delta-\kappa_t)\mathcal{E}$, then $ \langle \alpha_R\rangle=2\kappa\mathcal{E}/(i\Delta-\kappa_t)$ and $\langle \alpha_L\rangle=0$. In the following, we will simply assume that $|G_L|$ is suppressed either by a large detuning or by  adding a reverse pumping field to cancel the coupling exactly.

\begin{figure}[h]
\center \includegraphics[width=0.8\textwidth]{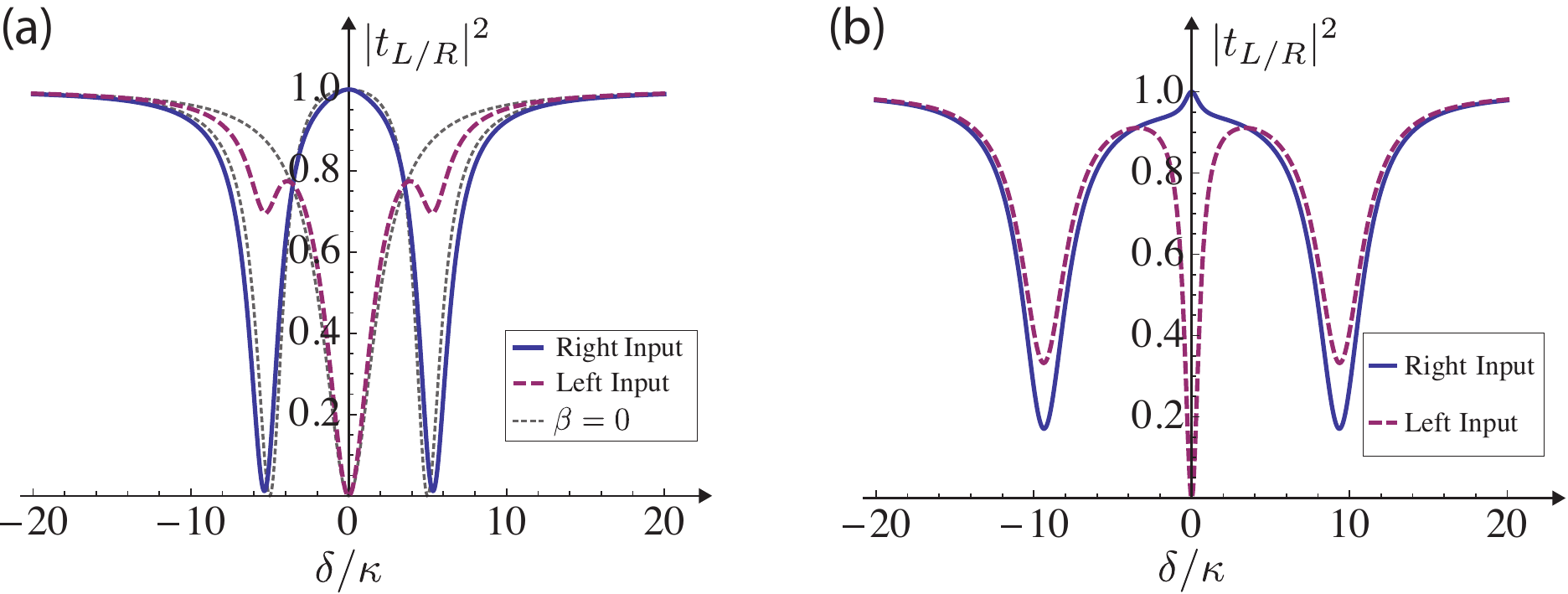}
\caption{ Transmittance for light propagating in a waveguide coupled to a resonator (AFP), in the presence of (a) weak ($\beta=2\kappa$)  and (b) strong ($\beta=8\kappa$) mode mixing. For these plots we have assumed $(\omega_{m},G_R,\kappa_{in},\gamma_m)/\kappa=(20,5,1,0)$ and $\Delta=-\omega_m$. \label{fig:AFP_mode_coupling}}
\end{figure}

In addition to pump backscattering, the probe photons are also mixed by the coupling term $\sim \beta$ in Hamiltonian (Eq.~(\ref{eq:hamiltonian})) and even for $|G_L|\rightarrow 0$ a degradation of  the non-reciprocal response of the device will occur. Let us first consider the case of weak mode mixing, $\beta\ll \kappa$, and  assume that  the system is pumped in the
right-circulating mode at the OM red sideband ($\Delta=-\omega_m, \omega_m\gg \beta$), as indicated  in Fig. \ref{fig:mode-coupling}(b). In this regime, the rate of backscattering of photons inside the resonator is smaller than the decay rate, and therefore, the non-reciprocal response of the device is qualitatively the same as in the ideal case. This is  shown in Fig. \ref{fig:AFP_mode_coupling}(a) where the mode coupling  only slightly reduces the operational bandwidth, i.e., $\kappa\rightarrow\kappa(1-\beta^2/G_R^2)$. 

\begin{figure}[h]
\center
\includegraphics[width=0.5\textwidth]{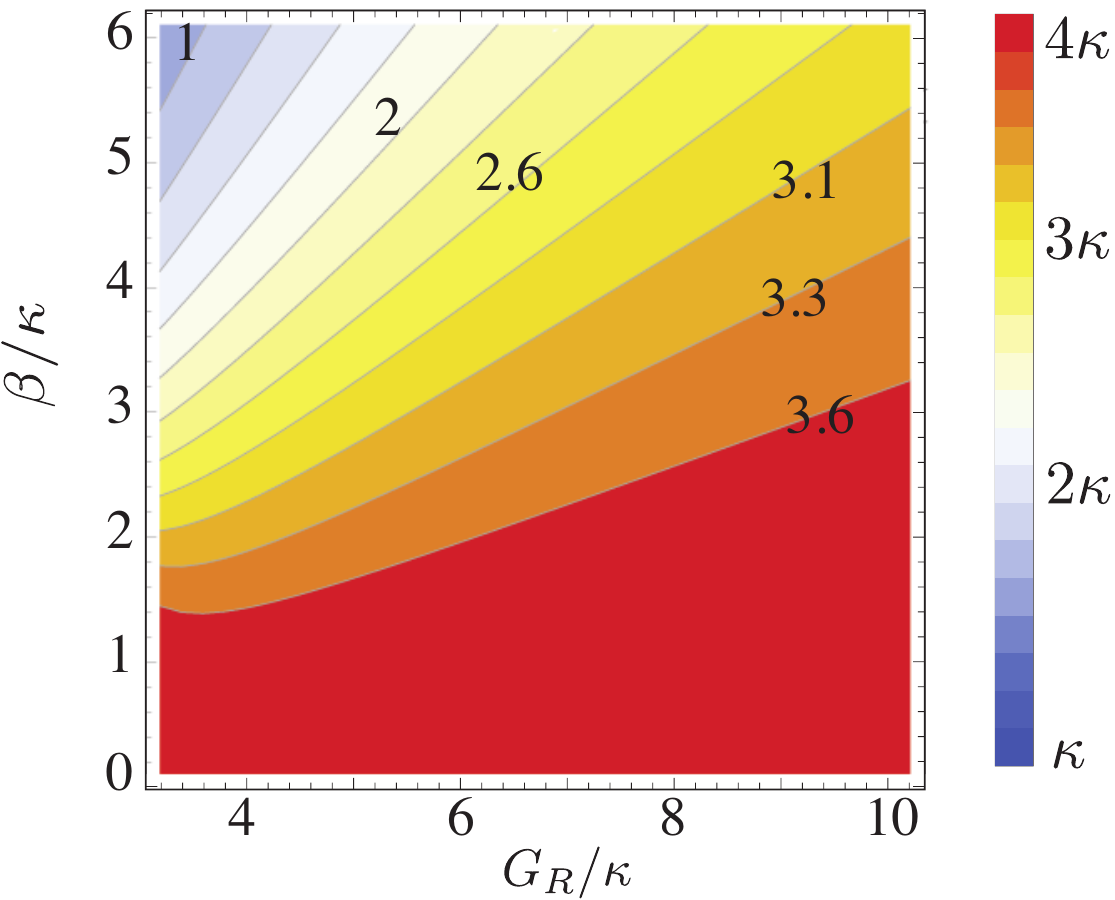}
\caption{ Operational bandwidth of an optical diode in the presence of a finite mode coupling $\beta$ and different values of the enhanced OM coupling $G_R$. For this plot we have assumed $G_L=0$ and  $(\omega_{m},\kappa_{in},\gamma_m)/\kappa=(20,1,0)$, $\Delta=-\omega_m$. In the absence of the mode coupling the bandwidth is $4\kappa$, which for a finite $\beta$ can be recovered by using a strong pump to enhance $|G_R|$. \label{fig:contrast_contour}}
\end{figure}

In contrast, when the mode coupling is strong $(\beta\gg \kappa)$,
the backscattering strongly redistributes the probe field in between right- and left-circulating
modes, and as shown in Fig. \ref{fig:AFP_mode_coupling}(b), the EIT width and the associated non-reciprocal effects can be significantly reduced. In Fig. \ref{fig:contrast_contour},  we have plotted the bandwidth of an optical diode as a function of the mode mixing and the strength of the OM coupling $|G_R|$. While the bandwidth decrease with increasing $\beta$, we observe that this effect can be compensated for by using a stronger pump to achieve $G_R>\beta$. Therefore,  we conclude that the presence of a finite intrinsic mode mixing does not fundamentally limit the operation of our device, and even if this coupling exceeds the ring resonator linewidth, non-reciprocal effects can persist, provided that the OM coupling is sufficiently strong.    

To put our results in relation with existing experimental parameters, we consider  the system presented in Ref. \cite{Verhagen:2011}, where  an optical  whispering gallery mode inside a toroidal microresonator is coupled to a mechanical mode of frequency $\omega_m/(2\pi)=78$ MHz. In this system the single-photon OM coupling is $g_0/(2\pi) = 3.4$ kHz and the directional enhanced coupling can reach $ G/(2\pi)= 11.4$ MHz. The resonator decay rate is  $\kappa_{t}/(2\pi)= 7.1$ MHz. Therefore, this system can be operated in the strong coupling regime $|G|>\kappa_t$, and assuming that intrinsic defects can be reduced to a level $|\beta| < |G|\sim 10$ MHz, this device can be used for implementing the different non-reciprocal effects described in this work.  In particular, if $\kappa\simeq\kappa_{in}$, then the optical isolation can be observed within the resonator bandwidth.
Note that recents experiments have demonstrated  OM systems supporting optical whispering  gallery modes with  mechanical frequencies  $\omega_m\sim$ GHz \cite{Carmon:2007,Ding:2010}. A further optimization of such devices could be used to achieve non-reciprocal OM effects at a much higher frequencies and to push the operational bandwidth of such devices into the 100 MHz regime.

\section{Thermal noise and the single photon limit}
So far we have only considered the scattering relations between the optical in- and out-fields, which due to the linearity of the equations of motion are the same for large classical fields as well as single photons. In practice additional noise sources will limit the operation of the device to a minimal power level, or equivalently to a minimal number of photons in the probe beams. A fundamental noise source in our system stems from the thermal Langevin force $\xi(t)$ which excites the mechanical resonator. The OM coupling up-converts mechanical excitations into optical photons which then appear as noise in the output fields \cite{Stannigel:2010,Chang:2010,Stannigel:2011}. To estimate the effect
of this noise, we investigate the contribution of thermal phonons in the noise power of the right moving out-field
\begin{equation}
P_{\rm noise}= \hbar \omega_c \times \int_{B} \frac{d\omega}{2\pi}\,  \langle \tilde{f}_{R,out}^{\dag1} (\omega) \tilde{f}_{R,out}^1(\omega) \rangle,
\end{equation}
where $B$ denotes frequency band of interest centered around the optical resonance.  We can use Eq.~(\ref{eq:full_output})  to express  $\tilde{f}_{R,out}^1 (\omega)$ in terms of the noise operator $\xi(\omega)$ and under the relevant conditions and $\Delta=-\omega_m$, we obtain the approximate result
\begin{equation}
P_{\rm noise}\simeq  \hbar \omega_c  \int_{B}  \frac{d\omega}{2\pi} \, \frac{ 2\gamma_m N_{th} \kappa G_R^2}{ G_R^4\!-\!2G_R^2(\omega\!-\!\omega_m)^2+ (\kappa_t^2+ (\omega\!-\!\omega_m)^2)(\omega\!-\!\omega_m)^2}.
\end{equation}
As described above, non-reciprocal effects are most effective in a small band around the mechanical frequency and we can set  $B=[\omega_m-\Delta B,\omega_m+\Delta B]$ where $\Delta B\ll \omega_m$ is the operation bandwidth of the device. By assuming that  $\Delta B\leq G_R^2/\kappa  $ in the weak coupling regime and $\Delta B\leq \kappa$ for strong OM coupling we obtain -- up to a numerical factor $\mathcal{O}(1) $ -- the general relation
\begin{equation}\label{eq:Pnoise}
P_{\rm noise}\approx  \hbar \omega_c \times \gamma_m N_{th}   \times \frac{\kappa \Delta B}{G_R^2}.
\end{equation}
For weak coupling and a maximal bandwidth $\Delta B= G_R^2/\kappa $, the noise power is given by the rate  $\gamma_m N_{th}\simeq k_BT/(\hbar Q_m)$ at which phonons in the mechanical resonator are excited. This means, that if we send a signal pulse of length $\Delta B^{-1}$ through the device a number $N_{\rm noise}\approx \gamma_m N_{th}  /(G_R^2/\kappa) $ noise photons is  generated during this time. Therefore, in this case the condition for achieving non-reciprocal effects on a single photon level, i.e. $N_{\rm noise}<1$, is equivalent to OM ground state cooling \cite{WilsonRae:2007p27575,marquardt2007qtc}, which is achievable in a cryogenic environment \cite{Verhagen:2011,Chan:2011}. Eq.~(\ref{eq:Pnoise}) also shows that the thermal noise level can be further reduced in the strong coupling regime.  In this case the maximal operation bandwidth is $\Delta B=\kappa$ and the noise power is suppressed by an additional  factor $(\kappa/G_R)^2\ll1$. This is due to the fact that thermal noise is mainly produced at the two split mode frequencies $\omega_m \pm G_R$, while the non-reciprocal effects rely on the transparency window between those modes. Note that while OM cooling saturates at $G_R\approx\kappa$,  the noise suppression in our device can always be improved with increasing ratio $G_R/\kappa$, eventually limited by the onset of the OM instability at $G_R=\omega_m/2$.

\section{Conclusions and outlook}

In summary, we have shown that optomechanics can induce non-reciprocity in the optical domain. In particular, an optomechanical ring resonator coupled to a waveguide induces a non-reciprocal phase in the under-coupled regime ($\kappa_{in}\ll\kappa$) and forms an optical isolator in the critically coupled regime ($\kappa \simeq \kappa_{in}$).

From an application perspective, this system provides an optical isolator that can be integrated on-chip. The bandwidth of such a device will be limited by the amount of pump power that the system can tolerate, before nonlinear effects become significant. In current experimental settings this amounts to bandwidths in the few MHz regime, which however could be further improved in optimized designs.

From a fundamental point of view, the relevant features of our technique are the possibility to implement coherent non-reciprocal phase shifts, to operate on the single photon level and the ability to dynamically control non-reciprocal effects by tuning the power of the pump beam. For example, one can consider a 2D array of optical resonators connected to each other via such non-reciprocal phase shifters. If the phase-shifts are chosen appropriately (e.g. according to the Landau gauge), then a tight-binding model of photons with an effective magnetic field can be simulated \cite{Hafezi:2011delay}. In other words, one can simulate quantum Hall physics with photons where the time-reversal symmetry is broken. In future experiments, it might be possible to combine these techniques with single photon non-linearities which could be either induced by the intrinsic non-linearity of the OM interaction itself \cite{Rabl:2011} or by interfacing the OM system with other atomic \cite{Aoki:2006p30830} or solid state qubits \cite{Stannigel:2010}.  Combined with such strong interaction between photons, the  implementation of magnetic Hamiltonians using micron-sized OM elements could pave the way for the exploration of fractional quantum Hall physics \cite{prange:Book,LesHouchesHall:BOOK} and various other exotic states of light.

\appendix
\section{Appendix: Phase sensitive transmission effects\label{sec:squeezing}}

\begin{figure}[h]
\center
\includegraphics[width=0.6\textwidth]{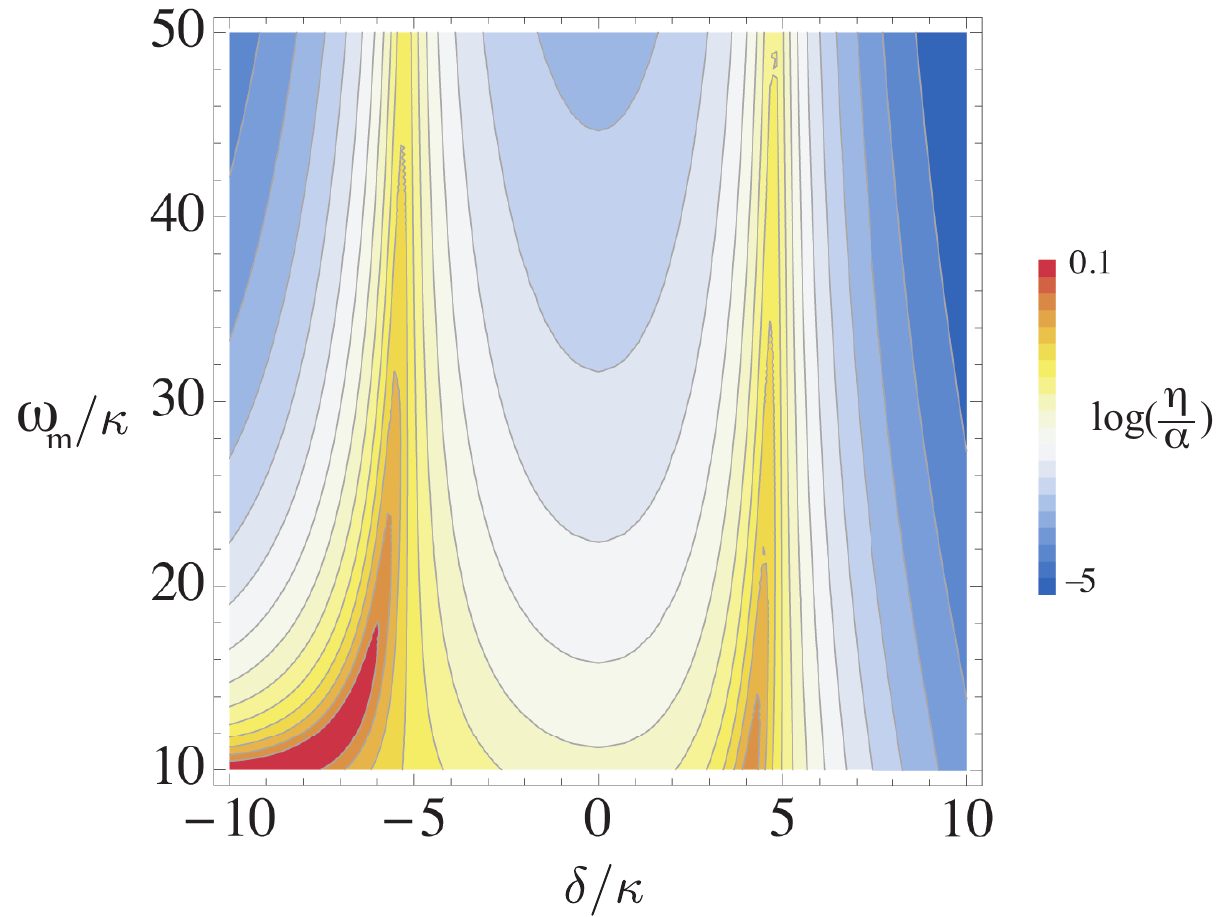}
\caption{Ratio between phase sensitive squeezing terms $(\eta)$ and the phase-insensitive transmission amplitudes $(\alpha)$.  For this plot we have assumed $\Delta=-\omega_{m}$ and $(G,\kappa_{in},\gamma_m)/\kappa=(5,.5,0)$.\label{fig:squeezing}}
\end{figure}

We study the effect of off-resonant OM interactions $\sim (a_i^\dag b^\dag + a_i b)$, which lead to a phase sensitive transmission or equivalently, a partial
squeezing of light in the output field. For simplicity, we consider \textcolor{black}{the case of a resonator coupled to a single waveguide} without mode coupling ($\beta=0$), assume $G_{R}=G$ and $G_L=0$, and set the pump field to the red sideband $(\Delta=-\omega_{m})$. In this situation, the
incoming right (left)-going field will exit the system as only right
(left)-going field, respectively. The out-field in the right-going channel is then given by
\begin{eqnarray}
\left(\begin{array}{c}
f_{R,out}^{1}(\omega)\\
f_{R,out}^{1\dagger}(-\omega)\end{array}\right) & = & \left(\begin{array}{cc}
\alpha(\omega) & \eta(\omega)\\
\eta^{*}(-\omega) & \alpha^{*}(-\omega)\end{array}\right)\left(\begin{array}{c}
f_{R,in}^{1}(\omega)\\
f_{R,in}^{1\dagger}(-\omega)\end{array}\right),
\end{eqnarray}
where
\begin{equation}
\alpha(\omega)=\frac{4|G|^{2}\text{\ensuremath{\omega_{m}}}(\text{\ensuremath{\omega_{m}}}+i\kappa)-\left(\omega^{2}-\text{\ensuremath{\omega_{m}}}^{2}\right)\left((\omega+i\text{\ensuremath{\kappa_{in}}})^{2}-(\omega_{m}+i\kappa)^{2}\right)}{\left(4|G|^{2}\text{\ensuremath{\omega_{m}}}^{2}+\left(\omega^{2}-\text{\ensuremath{\omega_{m}}}^{2}\right)\text{ }\left((\kappa+\text{\ensuremath{\kappa_{in}}}-i\omega)^{2}+\text{\ensuremath{\omega_{m}}}^{2}\right)\right)},
\end{equation}
and
\begin{equation}
\eta(\omega)=\frac{4iG^{2}\kappa\omega_{m}}{\left(4|G|^{2}\text{\ensuremath{\omega_{m}}}^{2}+\left(\omega^{2}-\text{\ensuremath{\omega_{m}}}^{2}\right)\left((\kappa+\text{\ensuremath{\kappa_{in}}}-i\omega)^{2}+\text{\ensuremath{\omega_{m}}}^{2}\right)\right)}.\end{equation}
The diagonal elements are the phase insensitive transmissions amplitudes, which we have discussed in the main part of the paper and which are related to the resonant OM coupling terms. In general the presence of non-zero off-diagonal terms, $\eta(\omega) \neq 0$, mixes the $f_{in}$ and $ f_{in}^\dag$ components. This implies a different transmission for different quadratures of the probe light, an effect which is exploited for OM squeezing, but is unwanted in the present context. However, as shown in Fig.\ref{fig:squeezing}, these effects are strongly suppressed in the parameter regime of interest.  In particular squeezing effects are negligible within the transparency window $|\delta| <G$, where non-reciprocal effects are most pronounced.

\section*{Acknowledgments}
The authors gratefully thank  A. Safavi-Naini, K. Srinvasan, J. Taylor, K. Stannigel and M. Lukin  for fruitful discussions.
This research was supported by the U.S. Army Research Office MURI award W911NF0910406, NSF through the Physics Frontier Center at the Joint Quantum Institute, the EU Network  AQUTE and by the Austrian Science Fund (FWF): Y 591-N16.


\begin{thebibliography}{10}

\bibitem{photonics_book:2008}
R.~B. Wehrspohn, H.~S. Kitzerow, and K.~Busch.
\newblock { \em{Nanophotonic materials: photonic crystals, plasmonics, and
  metamaterials}},
\newblock (Wiley-VCH, 2008).

\bibitem{Matthews:2009}
J.~C.~F.~Matthews, A.~Politi, A.~Stefanov, and J.~L.~O'Brien,
\newblock ``Manipulation of multiphoton entanglement in waveguide quantum
  circuits,''
\newblock { Nat. Photonics} \textbf{3}, 346--350 (2009).

\bibitem{Obrien:2009p27731}
J.~L.~O'Brien, A.~Furusawa, and J.~Vu{\v c}kovi{\'c},
\newblock ``Photonic quantum technologies,''
\newblock { Nat. Photonics} \textbf{3}, 687--695 (2009).

\bibitem{Sansoni:2010}
L.~Sansoni, F.~Sciarrino, G.~Vallone, P.~Mataloni, A.~Crespi, R.~Ramponi, and R.~Osellame,
\newblock ``Polarization entangled state measurement on a chip,''
\newblock { Phys. Rev. Lett.} \textbf{105}, 200503 (2010).

\bibitem{Politi:2009p45024}
A.~Politi, J.~C.~F.~Matthews, and J.~L.~O'Brien,
\newblock ``Shor's quantum factoring algorithm on a photonic chip,''
\newblock { Science} \textbf{325}, 1221 (2009).

\bibitem{Angelakis:2007}
D.~G.~Angelakis, M.~F.~Santos, and S.~Bose,
\newblock ``Photon-blockade-induced mott transitions and xy spin models in
  coupled cavity arrays,''
\newblock { Phys. Rev. A} \textbf{76}, 31805 (2007).

\bibitem{Greentree:2006}
A.~D.~Greentree, C.~Tahan, J.~H.~Cole, and L.~C.~L. Hollenberg,
\newblock ``Quantum phase transitions of light,''
\newblock { Nat. Phys.} \textbf{2}, 856--861 (2006).

\bibitem{Hartmann:2006}
M.~J. Hartmann, F.~G.~S.~L. Brandao, and M.~B.~Plenio,
\newblock ``Strongly interacting polaritons in coupled arrays of cavities,''
\newblock { Nat. Phys.} \textbf{2}, 849--855  (2006).

\bibitem{Potton:2004}
R.~Potton,
\newblock ``Reciprocity in optics,''
\newblock { Rep.  Prog.  Phys.} \textbf{67}, 717--754 (2004).

\bibitem{Espinola:2004}
R.~L.~Espinola, T.~Izuhara, M.~C.~Tsai, R.~M.~Osgood Jr, and H.~D{\"o}tsch,
\newblock ``Magneto-optical nonreciprocal phase shift in
  garnet/silicon-on-insulator waveguides,''
\newblock { Opt. Lett.} \textbf{29}, 941--943 (2004).

\bibitem{Levy:2005}
M.~Levy, 
\newblock ``Nanomagnetic route to bias-magnet-free, on-chip faraday rotators,''
\newblock { J. Opt. Soc. Am. B} {\bf22}, 254--260 (2005).

\bibitem{Zaman:2007}
T.~R.~Zaman, X.~Guo, and R.~J.~Ram,
\newblock ``Faraday rotation in an InP waveguide,''
\newblock { App.~Phys.~Lett.}  {\bf 90}, 023514 (2007).

\bibitem{Yu:2009}
Z.~Yu and S.~Fan,
\newblock ``Complete optical isolation created by indirect interband photonic
  transitions,''
\newblock { Nat. Photonics} {\bf 3}, 91--94 (2009).

\bibitem{Kang:2011}
M.~S. Kang, A.~Butsch, and P.~St.~J.~Russell,
\newblock ``Reconfigurable light-driven opto-acoustic isolators in photonic
  crystal fibre,''
\newblock { Nat. Photonics}, {\bf5}, 549--553 (2011).

\bibitem{Feng:2011}
L.~Feng, M.~Ayache, J.~Huang, Y.~-L.~Xu, M.~-H.~Lu, Y.~-F.~Chen, Y.~Fainman, and
  A.~Scherer,
\newblock ``Nonreciprocal light propagation in a silicon photonic circuit,''
\newblock { Science} {\bf333}, 729--733 (2011).

\bibitem{Fan:2011} 
S.~Fan, R.~Baets, A. Petrov, Z. Yu, J.~D.~Joannopoulos, W.~Freude, A.~Melloni, M.~ Popovic, M.~Vanwolleghem, D.~Jalas, M.~Eich, M.~Krause, H.~Renner, E.~Brinkmeyer, and C.~R.~Doerr, "Comment on Nonreciprocal light propagation in a silicon photonic circuit,"\newblock { Science} {\bf335}, 38 (2011).

\bibitem{Soljacic:2003}
M.~Solja{\v c}i{\'c}, C.~Luo, J.~D.~Joannopoulos, and S.~Fan,
\newblock ``Nonlinear photonic crystal microdevices for optical integration,''
\newblock { Opt. Lett.} {\bf 28}, 637--639 (2004).

\bibitem{Gallo:2001}
K.~Gallo, G.~Assanto, K.~Parameswaran, and M.~Fejer,
\newblock ``All-optical diode in a periodically poled lithium niobate waveguide,''
\newblock { Appl. Phys. Lett.} {\bf 79}, 314--316 (2001).

\bibitem{Manipatruni:2009}
S.~Manipatruni, J.~Robinson, and M.~Lipson,
\newblock ``Optical nonreciprocity in optomechanical structures,''
\newblock { Phys. Rev. Lett.} {\bf 102}, 213903 (2009).

\bibitem{Koch:2010}
J.~Koch, A.~A Houck, K.~Le Hur, and S.~M.~Girvin,
\newblock ``Time-reversal symmetry breaking in circuit-QED based photon lattices,''
\newblock { Phys. Rev. A} {\bf 82}, 043811 (2010).

\bibitem{Wang:2008}
Z.~Wang, Y.~Chong, J.~D.~Joannopoulos, and M.~Solja{\v c}i{\'c},
\newblock ``Reflection-free one-way edge modes in a gyromagnetic photonic
  crystal,''
\newblock { Phys. Rev. Lett.} {\bf100},13905 (2008).

\bibitem{Wang:2009p16784}
Z.~Wang, Y.~Chong, J.~D.~Joannopoulos, and M.~Soljacic,
\newblock ``Observation of unidirectional backscattering-immune topological
  electromagnetic states,''
\newblock { Nature} {\bf 461},772--775 (2009).

\bibitem{Haldane:2008}
F.~Haldane and S.~Raghu,
\newblock ``Possible realization of directional optical waveguides in photonic
  crystals with broken time-reversal symmetry,''
\newblock { Phys. Rev. Lett.} {\bf100}, 13904 (2008).

\bibitem{Hafezi:2011delay}
M.~Hafezi, E.~A.~Demler, M.~D.~Lukin, and J.~M.~Taylor,
\newblock ``Robust optical delay lines with topological protection,''
\newblock { Nat. Phys.} {\bf 7}, 907--912 (2011).

\bibitem{Umucalilar:2011}
R.~O.~Umucalilar and I.~Carusotto,
\newblock ``Artificial gauge field for photons in coupled cavity arrays,''
\newblock {Phys. Rev. A} {\bf 84}, 043804 (2011).

\bibitem{Verhagen:2011}
E.~Verhagen, S.~Del{\'e}glise, S.~Weis, A.~Schliesser, and T.~J.~Kippenberg,
\newblock ``Quantum-coherent coupling of a mechanical oscillator to an optical
  cavity mode,''
\newblock { arXiv:1107.3761}  (2011).

\bibitem{Chan:2011}
J.~Chan, T.~P.~Mayer Alegre, A.~H.~Safavi-Naeini, J.~T.~Hill, A.~Krause,
  Simon Gr{\"o}blacher, Markus Aspelmeyer, and Oskar Painter,
\newblock ``Laser cooling of a nanomechanical oscillator into its quantum ground
  state,''
\newblock { Nature} {\bf478}, 89--92 (2011).

\bibitem{Carmon:2007}
T.~Carmon and K.~Vahala,
\newblock ``Modal spectroscopy of optoexcited vibrations of a micron-scale
  on-chip resonator at greater than 1 ghz frequency,''
\newblock { Phys. Rev. Lett.} {\bf 98},123901 (2007).

\bibitem{Ding:2010}
L.~Ding, C.~Baker, P.~Senellart, A.~Lemaitre, S.~Ducci,
  Giuseppe Leo, and Ivan Favero,
\newblock ``High frequency gaas nano-optomechanical disk resonator,''
\newblock { Phys. Rev. Lett.} {\bf105}, 263903 (2010).

\bibitem{Stannigel:2010}
K.~Stannigel, P.~Rabl, A.~S.~S{\o}rensen, P.~Zoller, and M.~Lukin,
\newblock ``Optomechanical transducers for long-distance quantum communication,''
\newblock { Phys. Rev. Lett.} {\bf105}, 220501 (2010).

\bibitem{Chang:2010}
D.~E.~Chang, A.H.~Safavi-Naeini, M.~Hafezi, and O.~Painter,
\newblock ``Slowing and stopping light using an optomechanical crystal array,''
\newblock { New J.  Phys.} {\bf13}, 023003 (2011).

\bibitem{Stannigel:2011}
K.~Stannigel, P.~Rabl, A.~S.~S{\o}rensen, M.~D.~Lukin, and P.~Zoller,
\newblock ``Optomechanical transducers for quantum information processing,''
\newblock { Phys. Rev. A} {\bf 84}, 042341 (2011).

\bibitem{Fabre:1994}
C.~Fabre, M.~Pinard, S.~Bourzeix, A.~Heidmann, E.~Giacobino, and S.~Reynaud,
\newblock ``Quantum-noise reduction using a cavity with a movable mirror,''
\newblock { Phys. Rev. A} {\bf49},1337--1343 (1994).

\bibitem{WilsonRae:2007p27575}
I.~Wilson-Rae, N.~Nooshi, W.~Zwerger, and T.~Kippenberg,
\newblock ``Theory of ground state cooling of a mechanical oscillator using
  dynamical backaction,''
\newblock { Phys. Rev. Lett.} {\bf 99}, 093901 (2007).

\bibitem{marquardt2007qtc}
F.~Marquardt, J.~P.~Chen, A.~A.~Clerk, and S.~M.~Girvin,
\newblock ``Quantum theory of cavity-assisted sideband cooling of mechanical
  motion,''
\newblock { Phys. Rev. Lett.} {\bf 99}, 93902 (2007).

\bibitem{Schliesser:2010}
A.~Schliesser and T.~J.~Kippenberg,
\newblock ``Cavity optomechanics with whispering-gallery mode optical
  micro-resonators,''
\newblock { Adv. At., Mol., Opt. Phys.}
  {\bf 58}, 207--323 (2010).

\bibitem{Gardiner:1985}
C.~W.~Gardiner and M.~J.~Collett,
\newblock ``Input and output in damped quantum systems: Quantum stochastic
  differential equations and the master equation,''
\newblock { Phys. Rev. A} {\bf31}, 3761--3774 (1985).

\bibitem{fleischhauer1}
M.~Fleischhauer, A.~Imamoglu, and J.~P.~Marangos,
\newblock ``Electromagnetically induced transparency: optics in coherent media,''
\newblock { Rev. Mod. Phys.}, {\bf77}, 633--673  (2005).

\bibitem{Agarwal:2010}
G.~S.~Agarwal and S.~Huang,
\newblock ``Electromagnetically induced transparency in mechanical effects of
  light,''
\newblock { Phys. Rev. A}, {\bf81}, 041803 (2010).

\bibitem{Weis:2010}
S.~Weis, R.~Riviere, S.~Deleglise, E.~Gavartin, O.~Arcizet, A.~Schliesser, and T.~J.~Kippenberg,
\newblock ``Optomechanically induced transparency,''
\newblock { Science}, {\bf330},1520--1523 (2010).

\bibitem{SafaviNaeini:2011}
A.~H.~Safavi-Naeini, T.~P.~M.~Alegre, J.~Chan, M.~Eichenfield, M.~Winger, Q.~Lin, J.~T.~Hill, D.~E.~Chang, and O.~Painter,
\newblock ``Electromagnetically induced transparency and slow light with
  optomechanics,''
\newblock { Nature} {\bf 472}, 69--73 (2011).

\bibitem{Kippenberg:2002}
T.~J.~Kippenberg, S.~M.~Spillane, and K.~J.~Vahala,
\newblock ``Modal coupling in traveling-wave resonators,''
\newblock { Opt. Lett.} {\bf 27},1669--1671 (2002).

\bibitem{Mazzei:2007}
A.~Mazzei, S.~G{\"o}tzinger, L.~de~S.~Menezes, G.~Zumofen, O.~Benson, and
  V.~Sandoghdar,
\newblock ``Controlled coupling of counterpropagating whispering-gallery modes by
  a single rayleigh scatterer: a classical problem in a quantum optical light,''
\newblock { Phys. Rev. Lett.} {\bf99}, 173603 (2007).

\bibitem{Mancini:1994}
S.~Mancini and P.~Tombesi,
\newblock ``Quantum noise reduction by radiation pressure,''
\newblock { Phys. Rev. A} {\bf49}, 4055--4065 (1994).

\bibitem{Brooks:2011}
D.~Brooks, T.~Botter, N.~Brahms, T.~Purdy, S.~Schreppler, and D.~Stamper-Kurn,
\newblock ``Ponderomotive light squeezing with atomic cavity optomechanics,''
\newblock { arXiv:1107.5609}  (2011).

\bibitem{Rabl:2011}
P.~Rabl,
\newblock ``Photon blockade effect in optomechanical systems,''
\newblock { Phys. Rev. Lett.} {\bf107}, 063601 (2011).

\bibitem{Aoki:2006p30830}
T.~Aoki, B.~Dayan, E.~Wilcut, W.~P.~Bowen, A.~S.~Parkins, T.~J.~Kippenberg, K.~J.~Vahala, and H.~J.~Kimble,
\newblock ``Observation of strong coupling between one atom and a monolithic
  microresonator,''
\newblock { Nature}, {\bf443}, 671 (2006).

\bibitem{prange:Book}
R.~E.~Prange, S.~M.~Girvin, and M.~E.~Cage.
\newblock { \em The Quantum Hall Effect}.
\newblock (Springer-Verlag, 1986).

\bibitem{LesHouchesHall:BOOK}
A.~Comtet, T.~Jolicoeur, S.~Ouvry, and F.~David, editors,
\newblock { \em The Quantum Hall Effect: Novel Excitations and Broken
  Symmetries}
\newblock (Spinger-Verlag, 2000).

\end{thebibliography}
\end{document}